\documentclass[aps,prb,twocolumn,superscriptaddress,longbibliography]{revtex4-1}
\usepackage[colorlinks=true,citecolor=blue]{hyperref} 
\usepackage{amsmath}
\usepackage{amssymb}
\usepackage{hyperref}
\usepackage{siunitx}
\usepackage{graphicx}
\usepackage{dcolumn}
\usepackage{bm}

\usepackage{graphicx}
\usepackage{tabularx}
\usepackage{bbold}

\usepackage[usenames,dvipsnames]{xcolor}

\begin{document}

\title{Adversarial Hamiltonian learning of quantum dots in a minimal Kitaev chain}

\author{Rouven Koch}
\affiliation{Department of Applied Physics, Aalto University, 02150 Espoo, Finland}

\author{David van Driel}
\affiliation{QuTech, Delft University of Technology, 2600 GA Delft, The Netherlands}
\affiliation{Kavli Institute of Nanoscience, Delft University of Technology, 2600 GA Delft, The Netherlands}

\author{Alberto Bordin}
\affiliation{QuTech, Delft University of Technology, 2600 GA Delft, The Netherlands}
\affiliation{Kavli Institute of Nanoscience, Delft University of Technology, 2600 GA Delft, The Netherlands}

\author{Jose L. Lado}
\affiliation{Department of Applied Physics, Aalto University, 02150 Espoo, Finland}

\author{Eliska Greplova}
\affiliation{Kavli Institute of Nanoscience, Delft University of Technology, 2600 GA Delft, The Netherlands}%

\date{\today}

\begin{abstract}

Determining Hamiltonian parameters from noisy experimental measurements is a key task for the control of experimental quantum systems. An experimental platform that recently emerged, and where knowledge of Hamiltonian parameters is crucial to fine-tune the system, is that of quantum dot-based Kitaev chains. In this work, we demonstrate an adversarial machine learning algorithm to determine the parameters of a quantum dot-based Kitaev chain. We train a convolutional conditional generative adversarial neural network (Conv-cGAN) with simulated differential conductance data and use the model to predict the parameters at which Majorana bound states are predicted to appear. In particular, the Conv-cGAN model facilitates a rapid, numerically efficient exploration of the phase diagram describing the transition between elastic co-tunneling and crossed Andreev reflection regimes. We verify the theoretical predictions of the model by applying it to experimentally measured conductance obtained from a minimal Kitaev chain consisting of two spin-polarized quantum dots coupled by a superconductor-semiconductor hybrid. Our model accurately predicts, with an average success probability of $97$\%, whether the measurement was taken in the elastic co-tunneling or crossed Andreev reflection-dominated regime. Our work constitutes a stepping stone towards fast, reliable parameter prediction for tuning quantum-dot systems into distinct Hamiltonian regimes. Ultimately, our results yield a strategy to support Kitaev chain tuning that is scalable to longer chains.

\end{abstract}

\maketitle

\section{\label{sec:intro} Introduction}

The engineering of topological quantum matter represents a critical step for bringing theoretical predictions of exotic physics to the lab. One of the paradigmatic topological modes are Majorana bound states, associated with topological superconducting orders\cite{Sato2017,Beenakker2013,RevModPhys.87.137}, and have been studied extensively due to their potential for quantum computing~\cite{alicea2011non, aasen2016milestones}. The practical realization of Majorana bound states has proven to be greatly challenging, which motivated the development of multiple experimental platforms for their realization including semiconductors\cite{Prada2020}, atomic chains\cite{Feldman2016}, van der Waals heterostructures\cite{Kezilebieke2020}, and heavy-fermion materials\cite{Jiao2020}. Majorana bound states are predicted to appear at the edges of a spinless p-wave superconductor in 1D, the so-called Kitaev chain\cite{Kitaev2001,Beenakker2013,Sato2017,RevModPhys.87.137}. A minimal version of such a Kitaev chain has been recently experimentally realized in the system of two quantum dots that are mutually coupled both via crossed Andreev reflection and elastic co-tunneling\cite{PhysRevB.86.134528,Dvir2023}. The relative strengths of these two couplings can be tuned to drive the system into the so-called sweet-spot, where the electron tunneling and Andreev reflection have the same strength\cite{Wang2022,bordin2022controlled,Sau2012}. An arrival into the sweet-spot regime gives rise to the mid-gap modes consistent with a "Poor Man's Majorana bound state"\cite{PhysRevB.86.134528}. Fine-tuning a polarized quantum dot chain simultaneously presents a significant experimental challenge and is one of the limiting factors for scaling this type of quantum experiments\cite{Dvir2023}.

Inferring the description of a quantum system from available data, a problem known as Hamiltonian learning, is a central problem in quantum systems\cite{PhysRevLett.102.187203,PhysRevA.80.022333,PhysRevA.98.032114,Hincks2018,Wang2017,Gebhart2023}.
Neural network-based algorithms have recently risen to prominence as a tool for Hamiltonian learning and parameter estimation from experimental data\cite{PhysRevX.8.021026,PhysRevX.8.031029,PhysRevLett.122.150606,PhysRevLett.122.020504,PhysRevA.105.023302,PhysRevResearch.4.033223,PhysRevResearch.1.033092,2022arXiv221207893K,PhysRevResearch.3.023246}. While machine learning algorithms are by far not a universal answer to parameter determination challenges in noisy quantum systems, their generalization properties and fast evaluation make them suitable candidates to address technical questions connected to quantum control and parameter estimation. Neural network algorithms have been successfully used in supervised settings for both Hamiltonian estimation\cite{PRXQuantum.2.020303,Gentile2021,Wang2017} as well as tuning challenges\cite{PhysRevApplied.13.054019,PhysRevApplied.13.054005,PhysRevA.93.012122,PhysRevApplied.17.024069}. More recently, generative models have been shown to be powerful simulation and parameter estimation tools in the context of generic many-body physics problems\cite{PhysRevB.103.014301,PhysRevResearch.4.033223}.

\begin{figure}[t]
    \centering
    \includegraphics[width=0.47\textwidth]{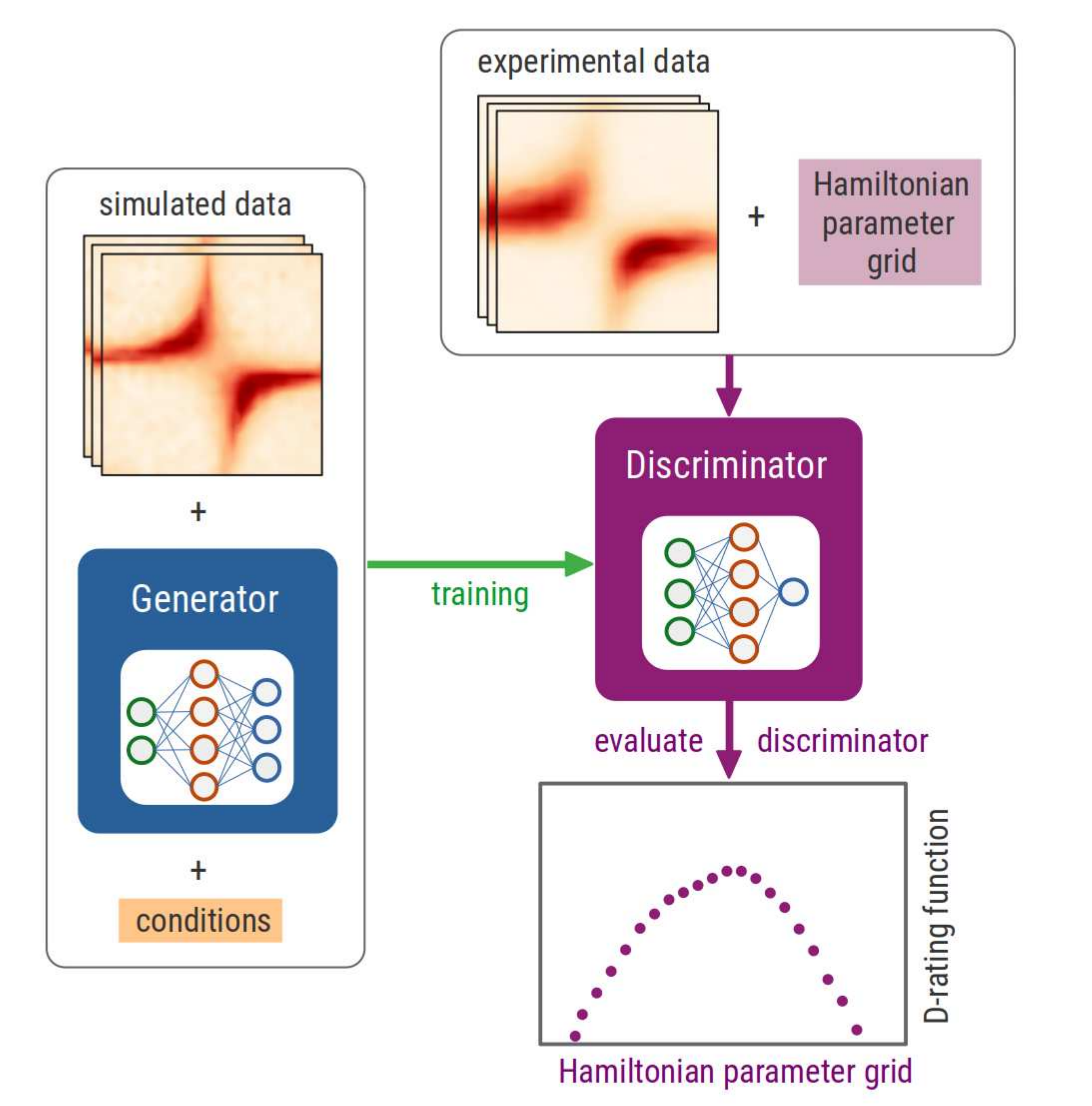}
    \caption{Workflow of the Hamiltonian learning process with the Conv-cGAN. The discriminator is trained with simulated data, the generator, and conditional parameters. The trained discriminator evaluates the experimental input on a discrete grid of values to obtain the Discriminator-rating function.}
    \label{fig:cGAN}
\end{figure}

Here, we demonstrate that convolutional conditional generative adversarial networks (Conv-cGAN) allow parameter estimation on both simulated and experimental data.
The key advantage of the conditional generative model is
the ability to extract information from (experimental) data and, more importantly, evaluate the ML output as a function of Hamiltonian parameters. We show that conditional generative models constitute a new figure of merit of parameter estimation on experimental data. We demonstrate our adversarial methodology in determining the Poor Man's Majorana sweet-spot from differential conductance measurement. We find that our model is able to produce full conductance spectra that are effectively indistinguishable from real measurement data and to estimate the ratio of crossed Andreev reflection and elastic co-tunneling rates with high precision.
Incidentally, our algorithm correctly identified measurement data instances that were later discovered to be mislabeled due to the large error bar in an experimental labeling procedure. Overall, the machine learning model reached $100$\% ($94$\%) accuracy on the first (second) measured data set in distinguishing tunneling- and Andreev reflection-dominated regimes. The two measured sets yielded average accuracy of $97$\% in regime recognition. The ability to quickly recognize the position with respect to the sweet-spot is key when driving the Kitaev chain to the fine-tuned point. Our algorithm thus presents a key stepping stone towards Majorana bound state tuning.

The work is organized as follows. In Section~\ref{sec:methods} we explain key methods used in this work: the physics of a Kitaev chain, the architecture and the training of conditional generative adversarial models, and details about experimental devices and the measurement procedure. In Section~\ref{sec:results} we summarize and explain the results we obtained on both simulated and experimental data and quantify the prediction power of our model. In Section~\ref{sec:conclusions} we summarize our results and outline the path of embedding pre-trained neural network models into the Majorana bound state tuning workflow.

\section{\label{sec:methods} Methods}

\begin{figure}[t]
    \centering
    \includegraphics[width=0.49\textwidth]{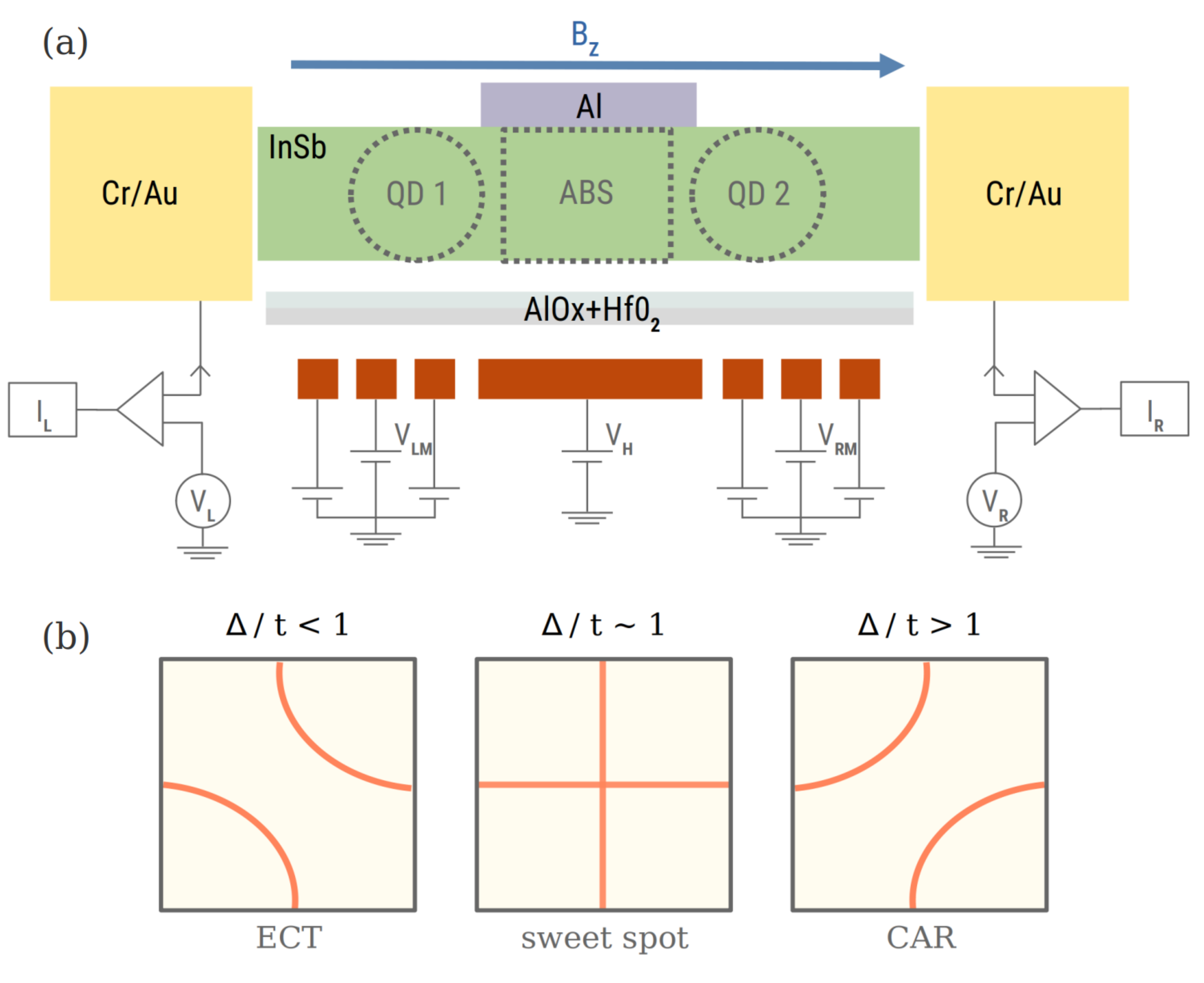}
    \caption{(a) Sketch of the experimental InSb nanowire setup containing two spin-polarized quantum dots (QD) and a hybrid hosting Andreev bound states (ABSs). Two Cr/Au normal leads allow for measuring the local and non-local differential conductance. More details about the device and measurement process are described in Sec.~\ref{sec:experiment}. (b) Sketch of the avoided crossings in the (local) differential conductance for the three different regimes of the Hamiltonian. At the sweet-spot (middle panel), the avoided crossings vanish.}
    \label{fig:intro}
\end{figure}

In this Section, we review the fundamental methodologies used in our manuscript. We first elaborate on the theoretical description of the Kitaev model and introduce the key parameters required for sweet-spot tuning. Afterwards, we provide a description of the Conv-cGAN architecture and the training process. 
We also summarize the key components of our experimental platform and provide details on the measurement process.

\subsection{Kitaev chain}

The model we consider to generate the theoretical training data is a minimal one-dimensional Kitaev model of two sites, modeling a 2-quantum-dot system with the Hamiltonian

\begin{equation}
    H = t d_L^\dagger d_R + \Delta d_L^\dagger d_R^\dagger + \epsilon_L d_L^{\dagger}d_L + \epsilon_R d_R^{\dagger}d_R + h.c.,
\end{equation}
where $t$ is the amplitude of elastic co-tunneling (ECT), $\Delta$ is the amplitude of crossed Andreev reflection (CAR), and $\epsilon_L$ ($\epsilon_R$) is the on-site energy of the left (right) quantum dot.
Here, ECT is the tunneling of a single electron from the left quantum dot to the right one. CAR is the reflection of an electron from the left quantum dot into a hole to the right quantum dot. Both are virtual tunneling processes mediated by the Andreev bound state (ABS) residing in the hybrid semiconducting-superconducting section~\cite{liu2022tunable,bordin2022controlled, Wang2022}.

The Hamiltonian can be written as
\begin{equation}
    H = \frac{1}{2} \Psi^\dagger h \Psi + \frac{1}{2} (\epsilon_L + \epsilon_R) \,.
\end{equation}
taking the Nambu spinor with the basis function $\Psi = (d_L,d_R,d_L^\dagger,d_R^\dagger)$ and the matrix $h$ has the form

\begin{equation}
    \label{Eq:h_matrix}
    h = 
    \begin{pmatrix}
        \epsilon_L & t & 0 & \Delta \\
        t & \epsilon_R & - \Delta & 0 \\
        0 & -\Delta & - \epsilon_L & -t \\
        \Delta & 0 & -t & -\epsilon_R 
    \end{pmatrix}
    \, .
\end{equation}

To emulate the experimentally accessible regime, we compute the differential conductance that accounts for the transport measurements through the double-dot system. Specifically, we consider each quantum dot to be tunnel-coupled to leads of normal metal with the rates $\Gamma_L$ and $\Gamma_R$ for the left and right dot, respectively. We will use the $\mathcal{S}$-matrix formalism, which describes the transmission and reflection coefficients of electrons and holes from the leads to each quantum dot and we can in turn relate these coefficients to the differential conductance. The $\mathcal{S}$-matrix can generally be written as
\begin{equation}
    \begin{split}
    S(\omega) &= 
    \begin{pmatrix}
        s_{ee} & s_{eh}  \\
        s_{he} & s_{hh}  
    \end{pmatrix} \\
    &=   \mathbb{1}  - i W^\dagger (\omega - H + \frac{1}{2} i WW^\dagger)^{-1} W \, ,
    \end{split} 
    \label{eq:smatrix}
\end{equation}
where $W=\text{diag}\{ \sqrt{\Gamma_L}, \sqrt{\Gamma_R}, -\sqrt{\Gamma_L}, -\sqrt{\Gamma_R} \}$ is the tunnel matrix with the tunnel coupling strengths $\Gamma_L$ and $\Gamma_R$ between the quantum dots and the normal leads. 

With the $\mathcal{S}$ matrix at hand, we compute the differential conductance, $G_{\alpha \beta}^0$ in the zero-temperature limit as
\begin{equation}
    G_{\alpha \beta}^0 = dI_\alpha / dV_\beta = \frac{e^2}{h} (\delta_{\alpha\beta} - | s_{ee}^{\alpha\beta}(\omega)|^2 + |s_{he}^{\alpha\beta}|^2 ) \, ,
\end{equation}
where $\alpha (\beta)$ defines the left or right quantum dot and $s_{ij}^{\alpha\beta}(\omega)$ refers to the matrix elements of the matrices $s_{ij}$ as defined in Eq.~\eqref{eq:smatrix}. Finally, to obtain the finite-temperature differential conductance we convolve the zero-temperature conductance with a Fermi-Dirac distribution. This leads to the expression
\begin{equation}
    G^{T}(\omega) = \int dE \frac{G^0(E)}{4k_BT\text{cosh}^2\left[(E-\omega)/2k_BT\right] } \, .
\end{equation}

In general, we can divide the parameter space and corresponding differential conductance into three regimes. In the two regimes where either ECT and CAR are the dominating effects, we obtain avoided crossings as depicted in Fig.~\ref{fig:cGAN}(b). In the "sweet-spot", where the magnitude of ECT and CAR is equivalent, the avoided crossings vanish. A more in-depth description of the processes behind the avoided crossings can be found in Ref.~\cite{Dvir2023}. Example experimental measurements of each regime can be found in Fig.~\ref{fig:fig5} (c,f,j).

We use the formalism described above to create the data for training the machine learning (ML) algorithm, described in the next section. For the data generation, we vary the parameters of Eq.~\ref{Eq:h_matrix} as well as the tunnel rates $\Gamma_L$ and $\Gamma_R$. In particular, we set the ECT magnitude $t=1$ and include $\Delta$-values between $0.3$ and $1.7$ to include a wide variety of parameters for the ECT-dominated regime ($t > \Delta$), the CAR regime ($t < \Delta$), as well as the sweet-spot ($t \approx \Delta$).\footnote{Setting the ECT magnitude $t=1$ is an approximation and one reason for differences compared to the experimental setup where $t$ varies across different measurements. We discuss this in more detail in the results section.}
The onsite energies $\epsilon_{L/R}$ are chosen randomly in the interval $L_\epsilon = [-1.5,1.5]$ around the center position $\epsilon_0$, which results in a shift of the center position of the avoided crossings. Additionally, we are adding background noise and a bias with the magnitude of $10^{-3}$. The coupling strengths to the leads $\Gamma_{L/R}$ are adjusted to match the experimental measurements.

\subsection{Conditional Generative Adversarial Networks}\label{sec:cGAN}

The key ingredient of our machine-learning approach to Hamiltonian learning from experimental data is a generative model conditioned on the experimentally tunable parameters. By training the generative model to reproduce the experimental measurements, we can study the dependence on successful and faithful data generation on these conditional parameters and thus determine the best match. Using a generative model in place of a feed-forward prediction has the additional advantage of visualizing the complete underlying parameter distribution and thus resulting in a more educated parameter estimate.

The ML algorithm we are using in this work for the Hamiltonian learning is a conditional generative adversarial network (cGAN)\cite{goodfellow2014generative,mirza2014conditional}. The general idea of GANs is to combine two separate deep neural networks (NNs) competing against each other in a min-max game during the training process. In the case of the cGAN, the generator network \textit{G} learns to map from a random input vector \textbf{z}, known as latent space, in combination with a conditional parameter \textbf{y} to the data distribution $p_{\text{data}}$ of input images of the training set. The discriminator network \textit{D} has the task to distinguish between samples of the training set and generated "fake" samples from the generator, given as constrains the conditional parameters.

The cGAN value function $V(\text{D},\text{G})$ is defined as

\begin{equation} 
    \begin{split}
    \label{Eq:val_cGAN}
        \text{min}_\text{G}\, \text{max}_\text{D} V(\text{D},\text{G}) & =
        \mathbb{E}_{x\thicksim p_{\text{data}}(\textbf{x})} [\text{log} \text{D}(\textbf{x}|\textbf{y})] \\
        & + \mathbb{E}_{z\thicksim p_z(\textbf{z})} [\text{log}~(1-\text{D}(\text{G}(\textbf{z}|\textbf{y}))],
    \end{split}
\end{equation}

where \textbf{x} is the input data, $p_z(\textbf{z})$ is the noise distribution of the latent space, and $\text{log}\left[1-\text{D}(\text{G}(z|y))\right]$ and $\text{log}[\text{D}(x|y)]$ are the corresponding terms for the generator and discriminator, respectively. In comparison of the original GAN value function\cite{goodfellow2014generative}, the additional conditional labels only appear in the terms corresponding to \textit{G} and \textit{D}.
During the training, the parameters of both networks get updated simultaneously by minimizing the generator term and maximizing the discriminator term. If trained successfully, the generator is able to create randomly new samples indistinguishable from the data distribution $p_{\text{data}}$. In contrast, the discriminator becomes very accurate in determining if a given input belongs to $p_{\text{data}}$ or is a "fake" sample.

Adding conditional Hamiltonian parameters to the GAN architecture allows us to control the physical parameters of the conductance maps created by the generator, which in the case of the original GAN would be purely random. In the following, we focus on the ratio between $\Delta$ and $t$ of the Hamiltonian of Eq.\eqref{Eq:val_cGAN} as the continuous conditional label which allows us to explore the complete phase space of the 2-site Kitaev chain. Due to the fact that we are dealing with input data with high spatial correlations among pixels, we choose an architecture for \textit{G} and \textit{D} consisting of convolutional neural networks (CNNs) (see App.~\ref{App:cGAN_network} for more details).

After the training process, both the generator and discriminator networks are trained equally well. For the Hamiltonian learning process, we exploit the trained discriminator to estimate parameters of a given measurement. It is worth noting that, the generation of conductance maps could be carried out with other generative algorithms such as diffusion models or autoencoders\cite{FlamShepherd2022}, but importantly, the existence of a discriminator in cGANs allows to directly use part of the trained architecture for Hamiltonian learning. In general, the discriminator provides the rating in the range between 0 and 1 of a given input array, with the corresponding label, to be from the real data set or the "fake" one.
To use the discriminator for the Hamiltonian learning process and to extract the most-likely parameter, we developed the following workflow, which can be divided into three basic steps which are also shown in Fig.~\ref{fig:cGAN}:
First, we define a landscape for the parameter range we want to estimate (here $\Delta/t \in I=[0.3,1.7]$).
Second, we feed the input spectrum, simulated or measured, into the discriminator in combination with every possible value of the parameter we want to extract.
Finally, the discriminator returns a score characterizing the likeliness that the input data provided corresponds to each possible parameter set. Combining all the outputs, we obtain a \textit{Discriminator-rating (D-rating) function}, estimating which Hamiltonian parameter is most likely belonging to the input image. In comparison to feed-forward NNs, this procedure gives us an uncertainty estimation of the maximum predicted value, and can naturally account for situations in which different parameter sets are consistent with the data.

\begin{figure}[t!]
    \centering
    \includegraphics[width=0.47\textwidth]{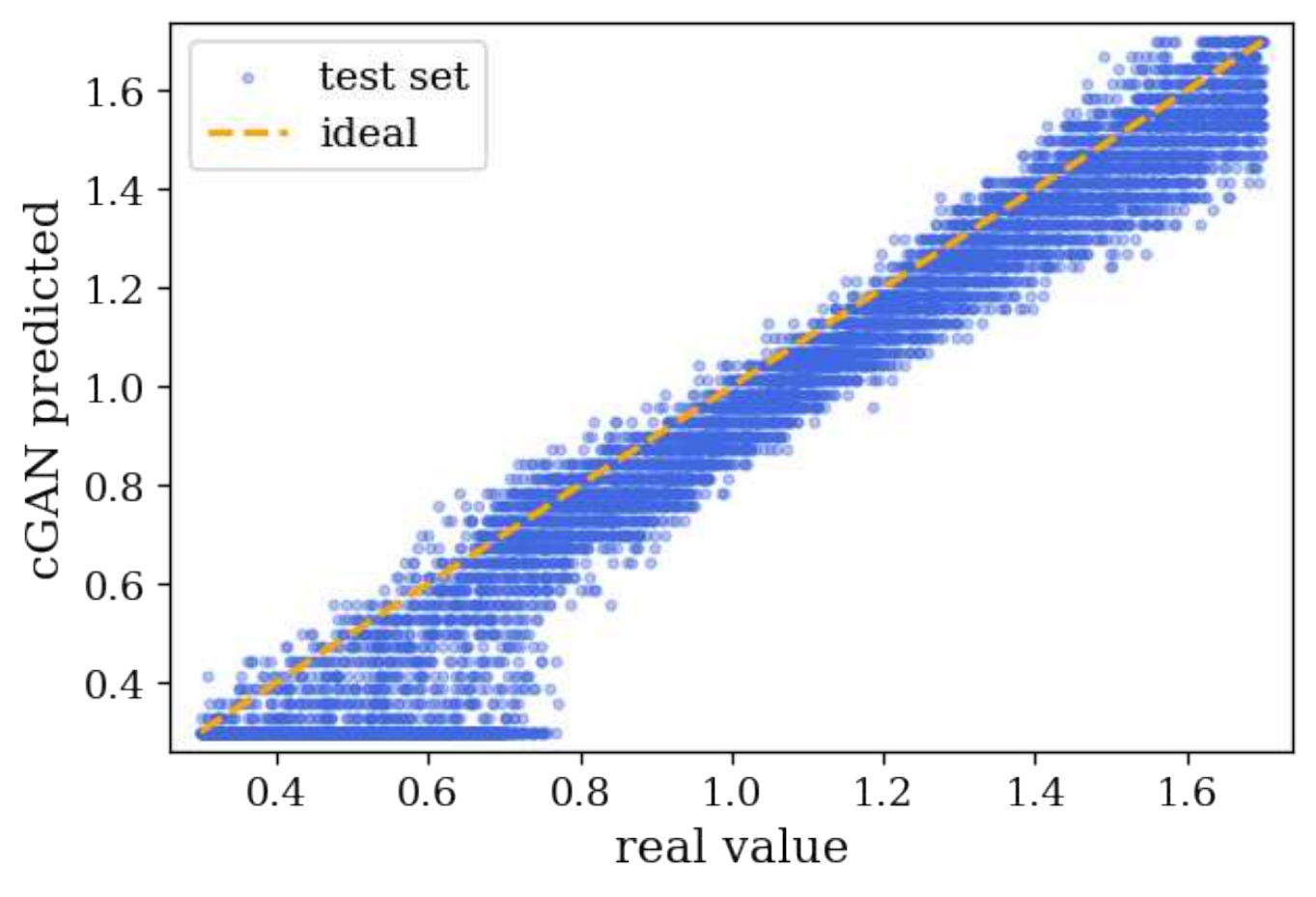}
    \caption{Conv-cGAN predictions vs. real values of the $\Delta/t$-ratio for the full test set consisting of 10.000 samples in the interval range from 0.3 to 1.7. The orange line shows the ideal prediction case.}
    \label{fig:results_kitaev_pred_vs_orig}
\end{figure}

\subsection{Experiments: 2-Quantum-dot system}
\label{sec:experiment}

The experimental data was obtained using the device sketched in Fig~\ref{fig:intro}. An InSb nanowire (NW) is placed on an array of bottom gates, which can change its electrochemical potential locally. 
Afterwards, the middle section of the NW is contacted by a thin ($\sim \SI{8}{\nano \meter}$) aluminum shell which is kept grounded throughout the experiment. The leftmost and rightmost three bottom gates are used to define quantum dots (QDs), whose electrochemical potentials are controlled by $V_\mathrm{LM}$/$V_\mathrm{RM}$ respectively. The NW is contacted by two normal, Cr/Au leads at its ends. These can be biased with a voltage $V_\mathrm{L/R}$ with respect to the grounded Al. For more details on device fabrication, see \cite{heedt2021shadow, Mazur.2022}. The middle gate $V_\mathrm{H}$ is used to tune the electrochemical potential of the hybrid InSb-Al section \cite{van2022electrostatic}. This changes the energy of Andreev bound states in the hybrid, which mediate ECT and CAR between the two QDs \cite{tsintzis2022creating}. As a result, ECT and CAR rates are modulated by $V_\mathrm{H}$ through their dependence on the ABS energy \cite{liu2022tunable, bordin2022controlled}.

\begin{figure}[t!]
    \centering
    \includegraphics[width=0.49\textwidth]{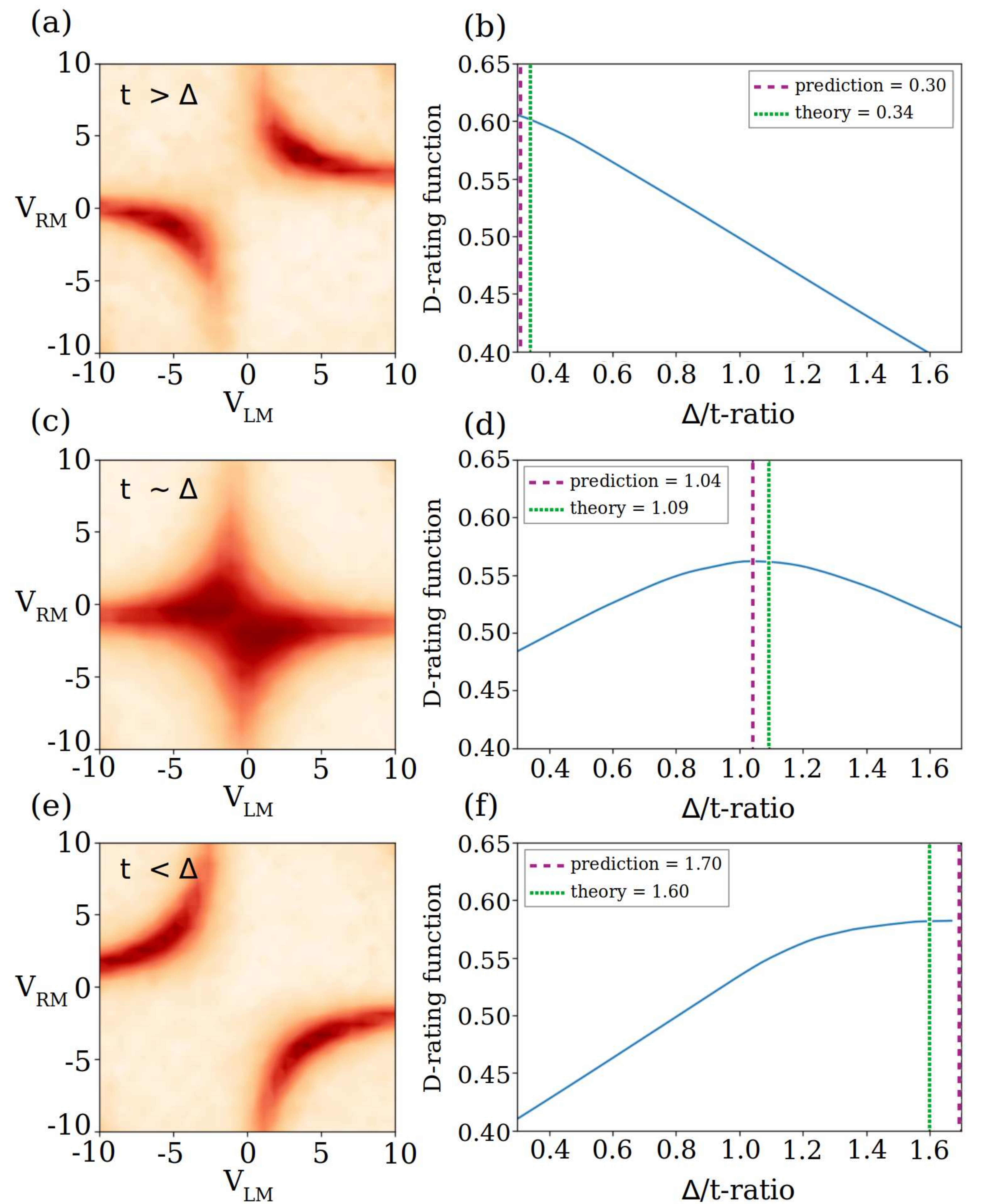}
    \caption{Predictions for the Kitaev model with the Conv-cGAN for the 3 regimes of the Hamiltonian. Examples are taken from the test set. (a,b) show an example for the $t>\Delta$ regime, (c,d) the "sweet-spot" around $t\approx\Delta$, and (e,f) the $t<\Delta$ regime. (a,c,e) are plots of the differential conductance $G$, (b,d,f) show the D-rating function corresponding to each image.} 
    \label{fig:results_kitaev_examples}
\end{figure}

The QD excitations are spin-polarized by applying an external magnetic field \cite{hanson2004semiconductor, van2022spin, danilenko2022spin}. We include in our analysis measurements for both $B_\mathrm{Z} = \SI{150}{\milli\tesla}$ and $B_\mathrm{Z} = \SI{100}{\milli\tesla}$ applied along the NW axis. The measurements to be classified by the cGAN are taken with both normal leads grounded ($V_\mathrm{L}=V_\mathrm{R}=0$).
The $\Delta/t$-ratio is determined using finite bias measurements. We set $V_\mathrm{L}=-V_\mathrm{R}=\SI{25}{\micro\electronvolt}$ to ensure only ECT occurs at a finite rate, which is maximum when $\epsilon_\mathrm{L}=\epsilon_\mathrm{R}$. In turn, we use a symmetric bias $V_\mathrm{L}=V_\mathrm{R}=\SI{25}{\micro\electronvolt}$ to have only CAR which is maximum when $\epsilon_\mathrm{L}=-\epsilon_\mathrm{R}$. The ratio between CAR and ECT can be estimated from finite bias experimental measurements through $\Delta/t \approx I_\mathrm{max}^\mathrm{CAR}/I_\mathrm{max}^\mathrm{ECT}$\cite{liu2022tunable}. These measurements are taken for different values of $V_\mathrm{H}$, together with $V_\mathrm{L}=V_\mathrm{R}=0$ QD-QD scans, to benchmark the cGAN predictions for different $\Delta/t$-ratio's. See supplementary Fig~\ref{fig:ECT_CAR_ratio_example} for an example determination of $I_\mathrm{max}^\mathrm{CAR}$, following the method of Wang et al \cite{wang2022triplet}.

\section{Results and Discussion}
\label{sec:results}
This section is divided into two parts. First, we focus on the predictions of the $\Delta/t$-ratio of the Conv-cGAN for the simulated data of the Kitaev model. Second, analyzing the results and predictions for the experimental measurements of the two-quantum-dot system in the InSb nanowire\cite{Dvir2023}.

\begin{figure}[t]
    \centering
    \includegraphics[width=0.5\textwidth]{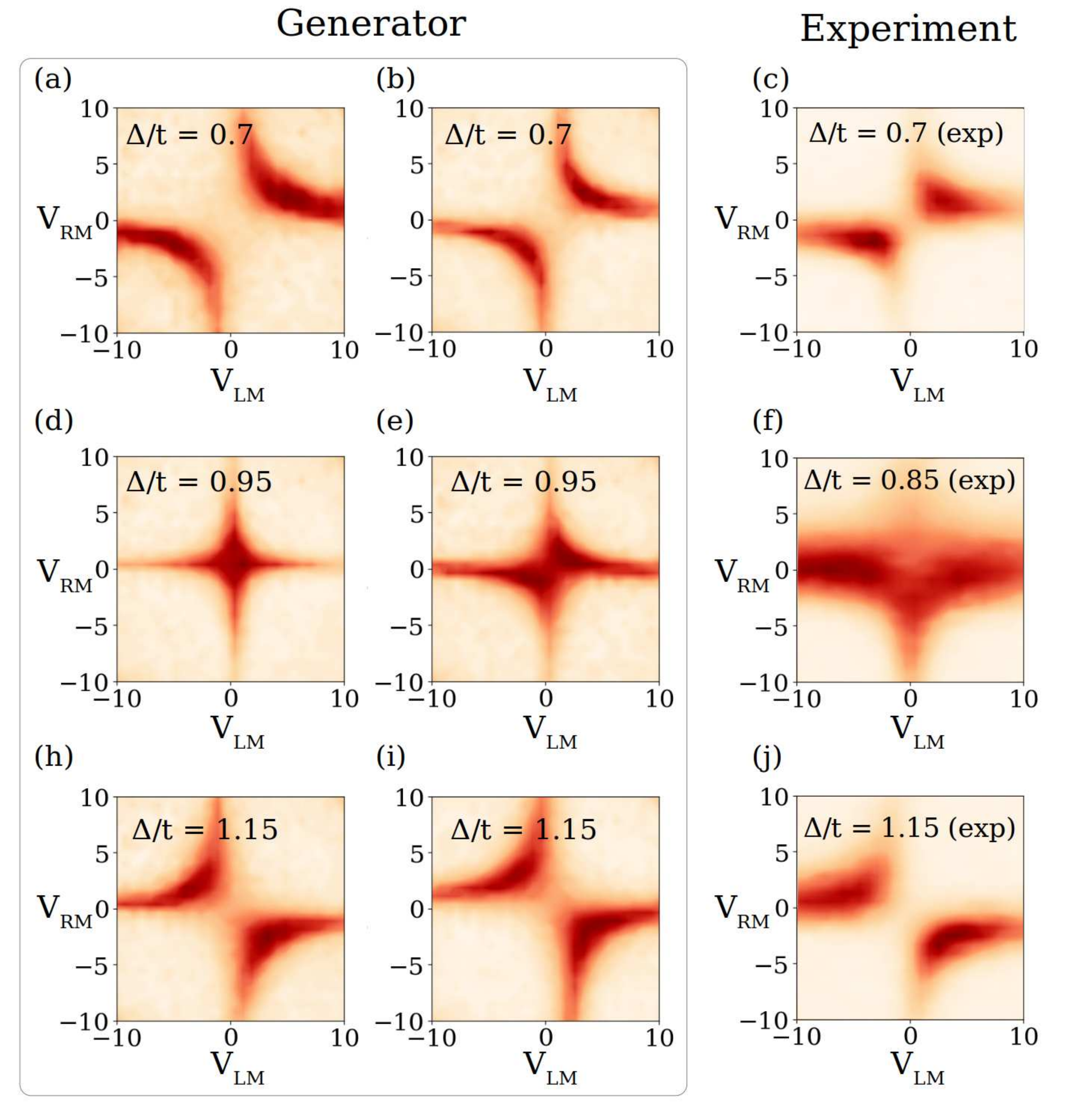}
    \caption{Generated images with the Conv-cGAN for the 3 parameter regimes in comparison to measurements. (a-c) show the ECT-dominated regime for $\Delta/t = 0.7$, (d-f) the vicinity of the sweet-spot for $\Delta/t = 0.95$ ($0.85$ from current measurements), and the CAR regime in (h-j) for $\Delta/t = 1.15$.
    The experimental images have been cut to 28$\times$28 pixels around the center of the avoided crossings.
    }
    \label{fig:generator_kitaev_examples}
    \label{fig:fig5}
\end{figure}

\subsection{Simulated data}
In this section, we apply the Conv-cGAN, in particular the trained discriminator, to predict the exact value of the $\Delta/t$-ratio of the simulated test set, consisting of 10000 samples. 
Figure~\ref{fig:results_kitaev_pred_vs_orig} shows the predicted versus real values of the whole test set. The training set consists of 140.000 samples for $\Delta/t \in [0.3, 1.7]$. The computed conductance images mimic the experimental measurements shown in the next section (more information about the training parameters can be found in App.~\ref{App:cGAN_network}). 
Taking all data points into account leads to a mean error ($\delta$) of $\delta_{test}=0.087$. In general, the predictions of the Conv-cGAN for the test set are very accurate and follow the ideal match (dotted orange line) closely. The deviations from the line are related to the intrinsic randomness of the cGAN algorithm as well as the added noise in the creation of the data set. The missing one-to-one correspondence of the $\Delta/t$ parameter for a given conductance image and the added noise leads to uncertainty when inferring the Hamiltonian. 
We observe in Fig.~\ref{fig:results_kitaev_pred_vs_orig} that the discriminator predicts the outer boundaries, especially $\Delta/t$=0.3, with higher probability than other values in the vicinity close to the boundaries. This might be related to an under-sampling of the boundary regions compared to parameters closer to the center.

Dividing the $\Delta/t$-ratio into the two regimes, the CAR and ECT regime above and below the sweet-spot, and using the Conv-cGAN as a classifier leads to an overall accuracy to predict the regime of 95.3$\%$ for the test set data. Taking into account that we predict the absolute values of $\Delta/t$, the parameter values directly around $\Delta/t=1.0$ lead to most of the classification errors and, therefore, the accuracy of 95$\%$ shows the predictive power of the Conv-cGAN. The classification is done by regression into larger or smaller $\Delta/t=1$ and then used to classify the regime, this leads to some errors around $\Delta/t=1$. If we add a "sweet-spot zone" between $0.95$ and $1.05$ we increase the accuracy to 97$\%$.

Figure~\ref{fig:results_kitaev_examples} shows three examples of conductance images, one for each parameter regime, randomly taken from the test set. The panels of Figure~\ref{fig:results_kitaev_examples}(a,c,e) show the generated conductance image with the corresponding D-rating function [panels Figure~\ref{fig:results_kitaev_examples} (b,d,f)], as explained in Sec.~\ref{sec:cGAN}. For all three cases, the D-rating function has a maximum at or in the near vicinity of the real maximum value. However, in the vicinity of the sweet-spot 
(Figure~\ref{fig:results_kitaev_examples}(c) and Figure~\ref{fig:results_kitaev_examples}(d)) the "D-rating function" does not decay as fast as for the other two cases with parameter values further away from the sweet-spot. Therefore, the discriminator is more confident in determining the maximum value in Figure~\ref{fig:results_kitaev_examples}(b) and Figure~\ref{fig:results_kitaev_examples}(f), showing a higher uncertainty for the predictions around the sweet-spot in (d). This uncertainty can not be determined with supervised NNs using CNNs which just predict an absolute value. The reason for the increased accuracy can be related to the blurred conductance image without a clear avoided crossing, which is characteristic when approaching the sweet-spot from both the ECT and CAR regime.

In addition to the Hamiltonian inference with the discriminator, the generator of the Conv-cGAN can generate high-quality conductance images for a given $\Delta/t$-ratio. In Fig.~\ref{fig:generator_kitaev_examples}, we used the generator to create images for $\Delta/t=0.7, 0.95, 1.15$ which can be compared with experimental measurements with the same or similar measured parameter shown in Fig.~\ref{fig:generator_kitaev_examples}(c,f,j).
The generated images for the ECT regime, illustrated in Fig.~\ref{fig:generator_kitaev_examples}(a,b), show the expected behavior of the avoided crossings as depicted in Fig.~\ref{fig:intro}(b). The differences between Fig.~\ref{fig:generator_kitaev_examples}(a) and Fig.~\ref{fig:generator_kitaev_examples}(b) are related to the intrinsic randomness of the Conv-cGAN which relates to noise and uncertainties in the experimental measurements as well as differences between the Kitaev model and real-world data.
The same effects can be seen in the sweet-spot regime in Fig.~\ref{fig:generator_kitaev_examples}(d) and Fig.~\ref{fig:generator_kitaev_examples}(e), as well as in the CAR regime in Fig.~\ref{fig:generator_kitaev_examples}(h) and Fig.~\ref{fig:generator_kitaev_examples}(i), where we obtain the expected avoided crossings with a similar magnitude and width.

In Fig.~\ref{fig:generator_kitaev_examples}, we observe good agreement with our simulations for the CAR regime ($\Delta/t$=1.15) when comparing \ref{fig:generator_kitaev_examples}(h,i) with \ref{fig:generator_kitaev_examples}(j). For the ECT-dominated regime (a-c), we observe differences in our model with the measurement \ref{fig:generator_kitaev_examples}(c) in the shape and size of the avoided crossings. The differences can be related to the approximation of $t=1=const$ which is usually not the case in the experimental setup where $t$ varies among different measurements which has an effect on the size and distance of the avoided crossings. 
For the experimental sweet-spot in \ref{fig:generator_kitaev_examples}(f) we picked the conductance image where the avoided crossings vanish and therefore visually can be determined as $\Delta/t \approx 1$. However, the current-inferred ratio is $\Delta/t = 0.85$ which hints at some difficulties in experimentally determining the $\Delta/t$-ratio. We find that by picking $\Delta/t$=0.95, the generator is able to reproduce the image of \ref{fig:generator_kitaev_examples}(f) accurately.

The individual analysis of the discriminator and generator in Figs.~\ref{fig:results_kitaev_examples} and~\ref{fig:generator_kitaev_examples} are a strong indication that the Conv-cGAN is trained properly. Furthermore, we want to highlight at this point again that the overall goal is to create an algorithm that works for both, theory as well as for experimental data, which will be analyzed in the next section.

\subsection{Experimental data}

The same Conv-cGAN used in the previous section is now applied to the local differential conductance measurements on the experimental two-quantum-dot system described in Sec.~\ref{sec:experiment}. We take the parameter-range of $\Delta/t \in [0.3, 1.7]$, covering a wide range of values. As explained in Sec.~\ref{sec:experiment}, changing the hybrid electrochemical potential with gate $V_\mathrm{H}$ leads to a wide variety of behavior in the conductance measurements of the quantum-dot system, ranging from the CAR to the ECT-dominated regime, including the sweet-spot condition of $\Delta\approx t$ where the "poor man's Majorana" modes are predicted to appear.
During the tuning process of the two-quantum-dot system into different Hamiltonian regimes, it is possible to experimentally extract values for the ratio $\Delta/t$ (see Sec.~\ref{sec:experiment}).

\begin{figure}[t!]
    \centering
    \includegraphics[width=0.5\textwidth]{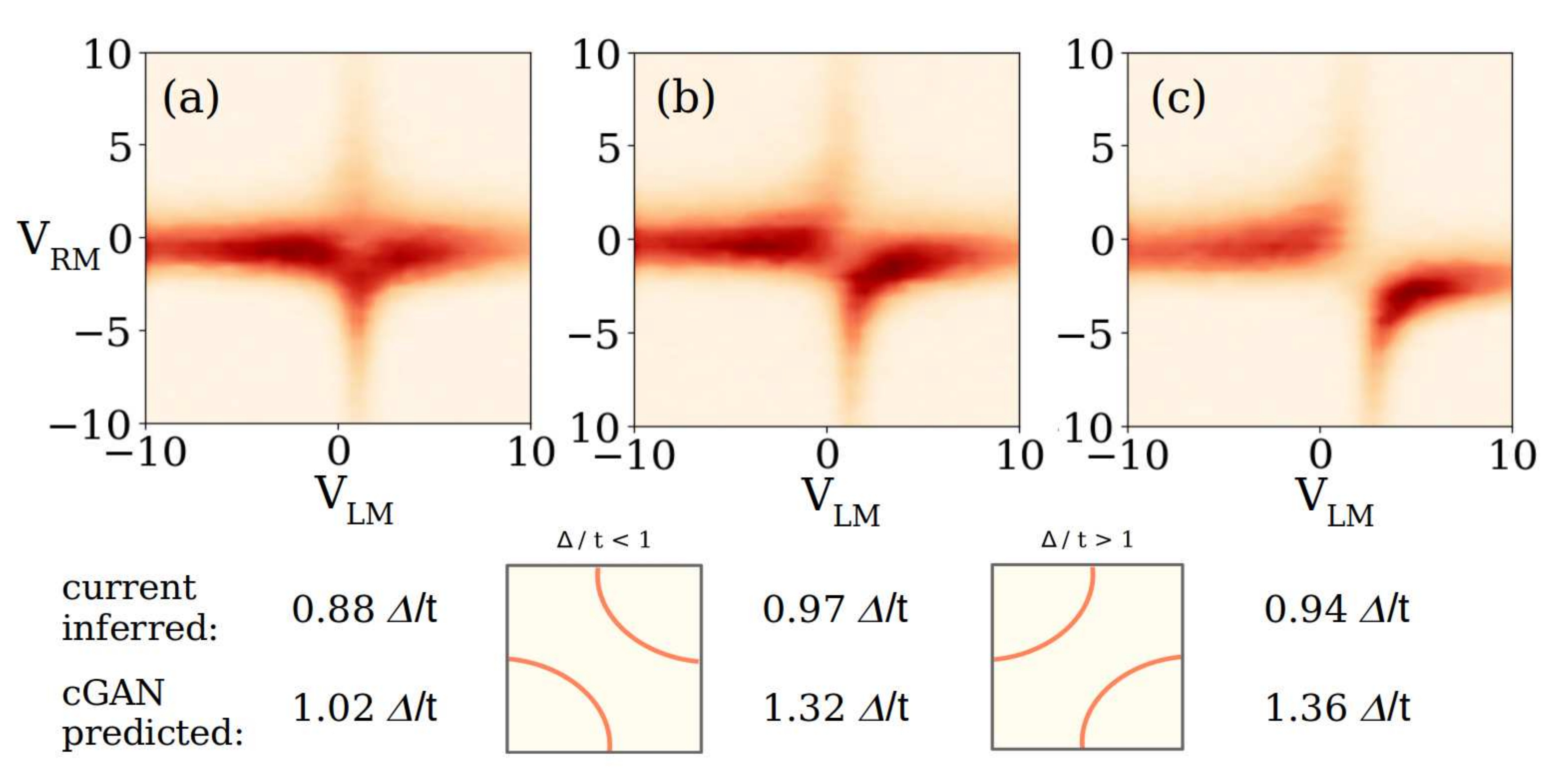}
    \caption{The three wrong-labeled measurements of Fig.~\ref{fig:wrong_label} with corresponding $\Delta/t$-ratio (a-c). The current-inferred parameter ratios of the measurements predict the avoided crossings visually to the wrong regime [see Fig.\ref{fig:cGAN}(b)]
    } 
    \label{fig:wrong_label}
\end{figure}

\begin{figure}[t!]
    \centering
    \includegraphics[width=0.5\textwidth]{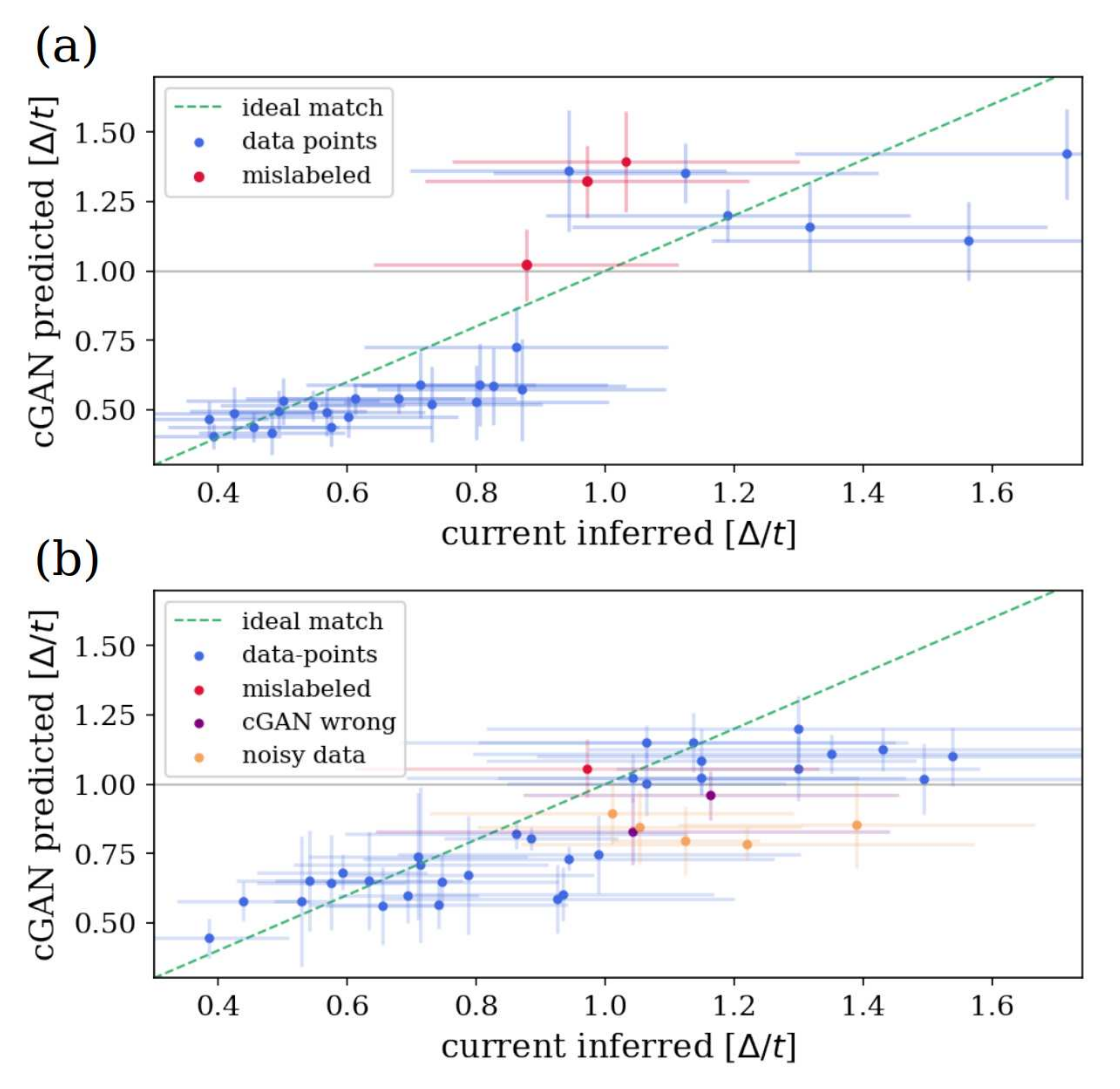}
    \caption{ Predictions of the Conv-cGAN vs. current-inferred labels of the $\Delta/t$-ratio for the two measurement series with a magnetic field of $B_\mathrm{Z} = \SI{150}{\milli\tesla}$ in (a) and $B_\mathrm{Z} = \SI{100}{\milli\tesla}$ in (b). The dotted green line shows the ideal match and the gray line marks the phase classification of the Conv-cGAN. The standard deviation of each measurement is given as an error bar in the x-y-direction and the prediction workflow is described in App.~\ref{App:workflow}.}
    \label{fig:predictions_cGAN}
\end{figure}

The experimental conductance measurements can be fed into the ML workflow as explained in App.~\ref{App:workflow}. In this work, we consider two conductance measurements series containing 29 and 39 measurements with the experimentally-extracted values for $\Delta/t \in [0.3, 1.7]$, measured at a magnetic field of $B_\mathrm{Z} = \SI{150}{\milli\tesla}$ and $B_\mathrm{Z} = \SI{100}{\milli\tesla}$, respectively.
The results of the 68 measurements are shown in Fig.~\ref{fig:predictions_cGAN}(a,b) and directly compare the Conv-cGAN predictions with the experimentally current-inferred values. 
We analyze both measurement series separately and show the (averaged) Conv-cGAN prediction vs. the current-inferred $\Delta/t$-ratio for 29 and 39 measurements. The standard deviation of the Conv-cGAN predictions are obtained via the workflow described in App.~\ref{App:workflow}, the error for the current-inferred parameters as described in Sec.~\ref{sec:experiment}.
Analyzing the 68 measurements, some of the current-inferred parameters in the parameter range between $\Delta/t=0.8$ and $\Delta/t=1.2$ show some inconsistencies between the current-inferred value and the appearance of the avoided crossings in the conductance measurements. 
For the first measurement series (Fig.~\ref{fig:predictions_cGAN}(a)), we detect three experimentally wrongly-labeled measurements, where the avoided crossings are visually in a different regime than determined by the current-inferred $\Delta/t$ value. These three mislabeled conductance measurements are shown in Fig.~\ref{fig:wrong_label} (a-c), in which the avoided crossings appear in the CAR-dominated regime of $\Delta/t > 1$ even though the measured parameters are smaller than 1, i.e. in the ECT regime. These measurements can visually be classified in the CAR. The predictions of the Conv-cGAN predict the correct regime for these three mislabeled measurements correctly. 
This is the first hint about the accuracy of the experimentally-extracted values and that our algorithm can also be used for outlier detection.
The averaged mean error between the measurement values and predictions of the Conv-cGAN, excluding the mislabeled measurements, is $\delta_{total_1}=0.164$ ($\delta_{total_1} = 0.148$) for the first 29 measurements. The classification into the correct regime leads to a classification accuracy of 89.7$\%$ taking all 29 measurements into account and 100$\%$ when neglecting the wrong labels shown in Fig.~\ref{fig:wrong_label}.
The predictions of the Conv-cGAN follow the trend of the "ideal match" with the experimentally current-inferred values well taking into account the huge error bars of the experimental values of $20-30\%$ and the cGAN uncertainties.
Especially for the ECT-dominated regime of $\Delta/t < 1$ the predictions follow the "ideal-matching" trend well opening the possibility for tuning the system towards the sweet-spot conditions.

Figure~\ref{fig:predictions_cGAN}(b) shows the results for the second measurement series, including 39 measurements at a Zeeman field of $B_\mathrm{Z} = \SI{100}{\milli\tesla}$. Similar to the first series, some difficulties arise around the sweet-spot of $\Delta/t \approx 1$. For five of the measurements, the noise levels are exceptionally high (orange dots) and the classification into the three different regimes becomes difficult since it is not possible for us to visually or experimentally verify the assigned label. Moreover, the Conv-cGAN predicts the wrong regime for two measurements (purple dots) and for one sample, the experimental current-inferred value is clearly in the wrong regime (red dot).
Excluding the noisy samples, we obtain a mean error of $\delta_{total_2} = 0.154$, whereas including the noisy samples leads to $\delta_{total_2} = 0.191$.
The accuracy to predict the correct regime is $94\%$ when we consider the 34 measurements, excluding the five noisy ones.

For the second measurement series, we observe that the Conv-cGAN predicted values are following the "ideal match" line with the current inferred parameter and we obtain similar results for the mean error and for the classification accuracy. Especially for the ECT-dominated regime ($\Delta/t < 1$) we clearly observe the linear trend towards the sweet-spot. 
In the vicinity of the sweet-spot ($\Delta/t \approx 1$) the magnitude of the errors increases, and therefore it is more difficult to make predictions or infer parameters close to the sweet-spot.
In the CAR-dominated regime ($\Delta/t > 1$) the upwards trend is not as clear as for the ECT regime and almost constant and the Conv-cGAN-predicted values are generally of smaller magnitude as the current-inferred values.

However, these results have to take into account different potential error sources: (i) general prediction errors of the Conv-cGAN discriminator due to the limited amount of training data and intrinsic randomness as discussed for Fig.~\ref{fig:results_kitaev_examples}, (ii) differences between the Kitaev model approach and real-world data considering the simplicity of the model as well as the approximations we used, and (iii) errors in determining the values experimentally by current measurements considering the huge error bars for these values.
As mentioned earlier, in the Kitaev Hamiltonian, we set $t=1$ and only vary $\Delta$ which is a simplified view compared to the experimental conditions where $t$ can vary during a measurement series. Therefore, by only considering the relation between $t$ and $\Delta$, our model does not consider differences in the avoided crossings related to the absolute magnitude of $t$. This could be a reason for the higher differences in the CAR-dominated regime in this measurement series.
Furthermore, in our analysis, we do not consider the "quality" of each individual measurement which is related to the noise and visibility of the avoided crossings. Images with very high noise magnitudes deviate from the theoretical model and the training data too much and, therefore, the Conv-cGAN can not make as accurate predictions about them as for "perfect" measurements. 
Considering the experimentally wrong-labeled measurements and relatively high error bars, the current-inferred values can not be taken as ground truth parameter, but rather be used to give us an idea about the accuracy of the Conv-cGAN predictions. For most of the mislabeled measurements, the Conv-cGAN predicts the correct parameter regime.

Another reason why the vicinity of the sweet-spot is harder to predict is that there are only very small visual differences on each side of the sweet-spot in the local conductance. The absence of clear avoided crossings and higher noise levels lead to higher uncertainties in the prediction. Therefore, to predict the sweet-spot regime accurately, very clean or additional measurements such as the non-local conductance are required.

Following the previous discussion, we do not expect that the Conv-cGAN predictions with the current-inferred extraction will yield an ideal match line since the experimental values can not be seen as perfect values. Furthermore, corrections beyond the minimal model considered can lead to deviations when applying the algorithm to real-world data.
Nonetheless, we observe an overall good agreement between our ML algorithm trained on the Kitaev model and the experimental data, the strength of our approach to classify the correct regime and detect outliers in the current-based parameter extraction.

\begin{figure}[t!]
    \centering
    \includegraphics[width=0.48\textwidth]{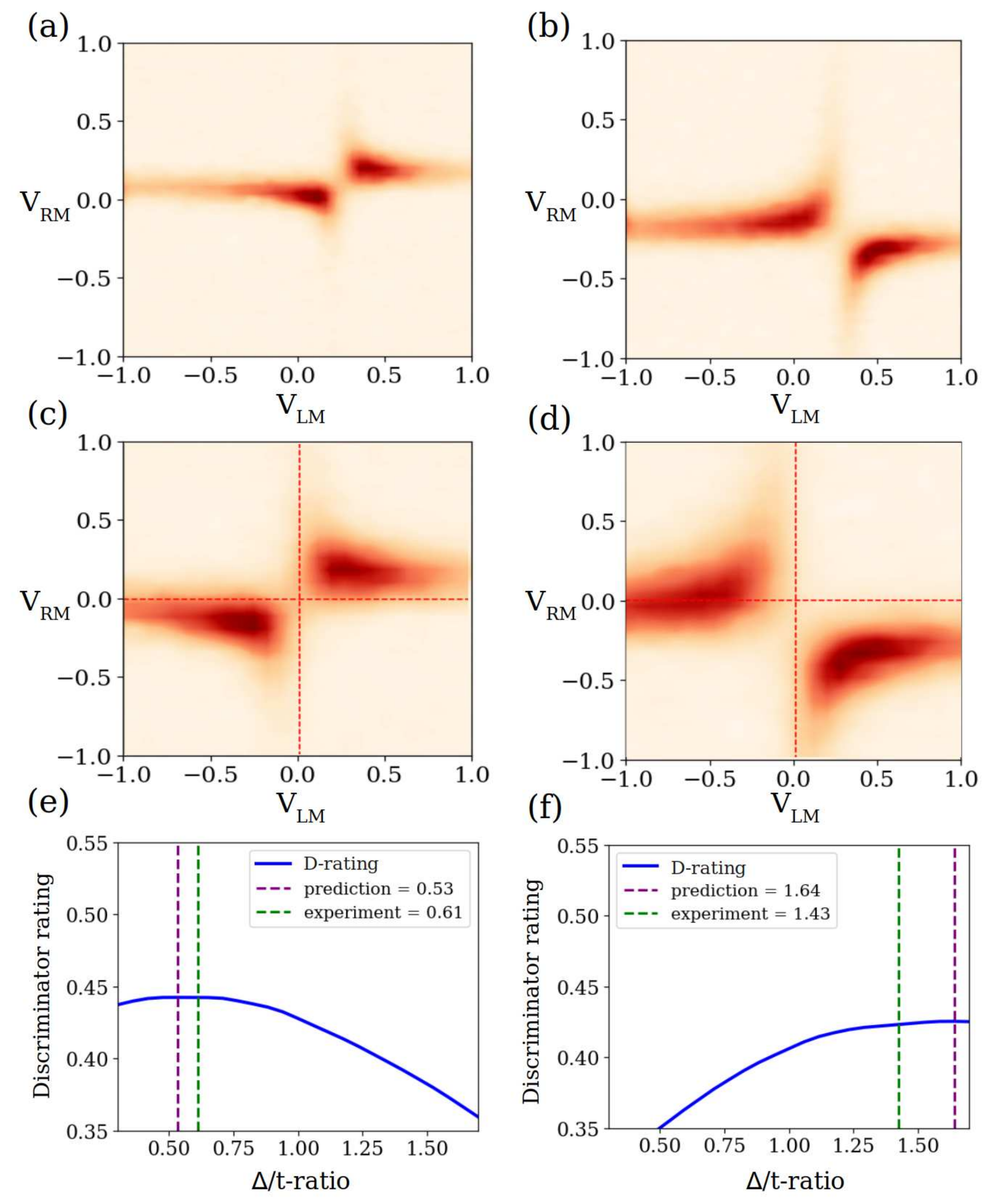}
    \caption{Predictions for the experimental measurements of the 2-quantum dot system with the Conv-cGAN trained on the full data set for two examples of the measurement series. The original measurements are shown in (a) and (b). (c) and (d) are the corresponding cropped images which are the inputs to the cGAN. (e) and (f) show the Discriminator-rating function predicted by the discriminator of the Conv-cGAN.}
    \label{fig:results_exp}
\end{figure}

The procedure of predicting the $\Delta/t$ value with the Conv-cGAN is illustrated in Fig.~\ref{fig:results_exp} for two example measurements. The measurement data, shown in Fig.~\ref{fig:results_exp}(a,b), is cropped to $28\times28$ pixels Fig.~\ref{fig:results_exp}(c,d) which is the input size of the Conv-cGAN. The cropped and normalized image is then fed into the discriminator leading to the D-rating function Fig.~\ref{fig:results_exp}(e,f) for each measurement. The cropping of the image can be done arbitrarily around the center which, however, leads to slightly different results. The averaging process to extract the Hamiltonian parameter for the results in Fig.~\ref{fig:predictions_cGAN} is explained in more detail in App.~\ref{App:workflow}.

Let us conclude this section with a benchmark comparison to a standard feed-forward mode. We used the same Kitaev chain model, described in Sec.~\ref{sec:methods}, and training data used for the Conv-cGAN to train a supervised CNN network and make predictions for the same 68 experimental data points. 
In Fig.~\ref{fig:predictions_CNN}, we show the CNN-predicted $\Delta/t$ value compared with the current-inferred value for the two measurement series at $B_\mathrm{Z} = \SI{150}{\milli\tesla}$ in Fig.~\ref{fig:predictions_CNN}(a) and $B_\mathrm{Z} = \SI{100}{\milli\tesla}$ in Fig.~\ref{fig:predictions_CNN}(b). 
In general, the classification accuracy of the CNN is 92 and 94$\%$ and, therefore, slightly lower compared to the Conv-cGAN value. The mean error for both measurement series is $\delta_{CNN-1}=0.150$ and $\delta_{CNN-2}=0.170$. Furthermore, the predictions for the ECT-dominated regime, are almost constant and not following the trend of the "ideal match" as it was the case before. More details can be found in App.~\ref{App:CNN}. In general, we find that the Conv-cGAN yields better results than the simple supervised-learning approach with CNNs when estimating the ECT/CAR amplitude ratio as well as for the classification of the regimes.

\section{Conclusion}
\label{sec:conclusions}

 In this work, we show the first demonstration of cGANs for Hamiltonian learning of simulated and real experimental measurements in quantum dot-based Kitaev chain systems that host Poor Man's Majoranas in a parameter sweet-spot. We show that our algorithm is capable of predicting both simulated and experimental measurements with good accuracy as well as giving an estimation about the uncertainty of each prediction via the discriminator-rating function. This is a clear advantage over supervised neural networks predicting just a single value for each input. This work can be seen as the first proof-of-principle that cGANs can be a powerful tool to support experimental measurements and extract underlying Hamiltonians from real-world problems. It is important to note that, taking into account all sources of error and the still existing difficulties in training GANs, there is room for improvements in the training of the GAN as well as pre-and post-processing of especially the experimental data.
 Interestingly, we show that even when training the Conv-GAN on the relatively simple theoretical model yields robust Hamiltonian prediction and in the case presented here outperforms previously known labeling methods for the experimental measurements. We thus put forward a new figure of merit for experimental tuning up Kitaev chain systems into the $t = \Delta$ sweet-spot.

\begin{acknowledgments}
R.K. and J.L.L. acknowledge
the computational resources provided by
the Aalto Science-IT project,
and the
financial support from the
Academy of Finland Projects Nos. 331342, 336243 and 358088,
and the Jane and Aatos Erkko Foundation. E.G. acknowledges project Engineered Topological Quantum Networks (with project number VI.Veni.212.278) of the research programme NWO Talent Programme Veni Science domain 2021 which is financed by the Dutch Research Council (NWO). E.G. and R.K. acknowledge support of the Kavli Institute of Nanoscience Delft.
\end{acknowledgments}

\section*{Code and Experimental Data Availability}
This paper is supplemented by a GitLab repository with all the code and data necessary to reproduce our results~\url{https:
//gitlab.com/QMAI/papers/adversarialkitaevchain}.
The raw experimental data, as well as code used for labeling, are available at \url{https://zenodo.org/record/7798010}.

\appendix

\section{Architecture and Training of the Conv-cGAN}
\label{App:cGAN_network}

\textbf{Architecture:}
The code for the Conv-cGAN and for the CNN can be found in Ref.~\cite{AdversarialKitaevChain}. The code uses Tensorflow~\cite{tensorflow2015} and Keras~\cite{chollet2015keras}.
The generator architecture of the Conv-cGAN is shown in Table~\ref{tab:generator}, consisting of 8 layers and in total 995329 (trainable) parameters. The layers include an input concatenation of the latent vector (dimension 100) with the continuous conditional parameter, a Dense layer followed by deconvolutional layers (Conv2DT) with Dropout of $25\%$ between deconvolutional layers. The output dimension is ($28,28,1$) since are using "gray-scale" images with the size of $28\times28$ pixels.

The discriminator network is shown in Table~\ref{tab:discriminator} and consists of 11 layers and 878721 total parameters. The input layer with dimension ($28,28,1$) is followed by two times the combination of a convolutional layer (Conv2D), a max pooling operation (i.e., down-sampling along its spatial dimensions), and a dropout layer. The result is flattened to a 1-dimensional array and concatenated with the conditional parameter, before being fed into two Dense layers.
We are using the ReLU activation function~\cite{relu2018} for hidden layers and a sigmoid function for the last layers. We are using the Adam optimizer~\cite{kingma2014adam} for the training of the total Conv-cGAN with a learning rate of 0.00001. For the Generator, we use the MSE as a loss metric, for discriminator binary cross-entropy.

\textbf{Training:}
The training set consists of 140000 conductance images, generated with the Kitaev model introduced in Sec.~\ref{sec:methods}. The batch size is set to 32 and we trained the network for a total number of 40 epochs. The training was performed on a Nvidia GeForce GTX 1070 GPU which takes approximately 5 hours.

\begin{table}[h]
    \centering
    \caption{Architecture of the generator network of the Conv-cGAN.}
    \label{tab:generator}
        \begin{tabularx}{.48\textwidth}{ |X X| }
          \hline
          \multicolumn{2}{|c|}{Generator} \\
          \hline
          \hline
          Layer & Output shape \\
          \hline
          Concatenation & 101 \\
          Dense & 6272 \\
          Reshape & (7,7,128) \\
          Conv2DT & (14,14,128) \\
          Dropout & (14,14,128) \\
          Conv2DT & (28,28,64) \\
          Dropout & (28,28,64) \\
          Conv2DT & (28,28,1) \\
          \hline
          total parameters & 995,329 \\
          \hline
        \end{tabularx}
\end{table}

\begin{table}[h]
    \centering
    \caption{Architecture of the discriminator network of the Conv-cGAN.}
    \label{tab:discriminator}
        \begin{tabularx}{.48\textwidth}{ |X X| }
          \hline
          \multicolumn{2}{|c|}{Discriminator} \\
          \hline
          \hline
          Layer & Output shape \\
          \hline
          Input & (28,28,1) \\
          Conv2D &  (28, 28, 64) \\
          MaxPooling2D & (14, 14, 64) \\
          Dropout & (14, 14, 64) \\
          Conv2D &  (14, 14, 128) \\
          MaxPooling2D & (7, 7, 128) \\
          Dropout & (7, 7, 128) \\
          Flatten & 6272 \\
          Concatenate & 6273 \\
          Dense & 100 \\
          Dense & 1\\
          \hline
          total parameters & 878,721 \\
          \hline
        \end{tabularx}
\end{table}

\section{Results: supervised CNN benchmark}
\label{App:CNN}

\begin{figure}[t!]
    \centering
    \includegraphics[width=0.49\textwidth]{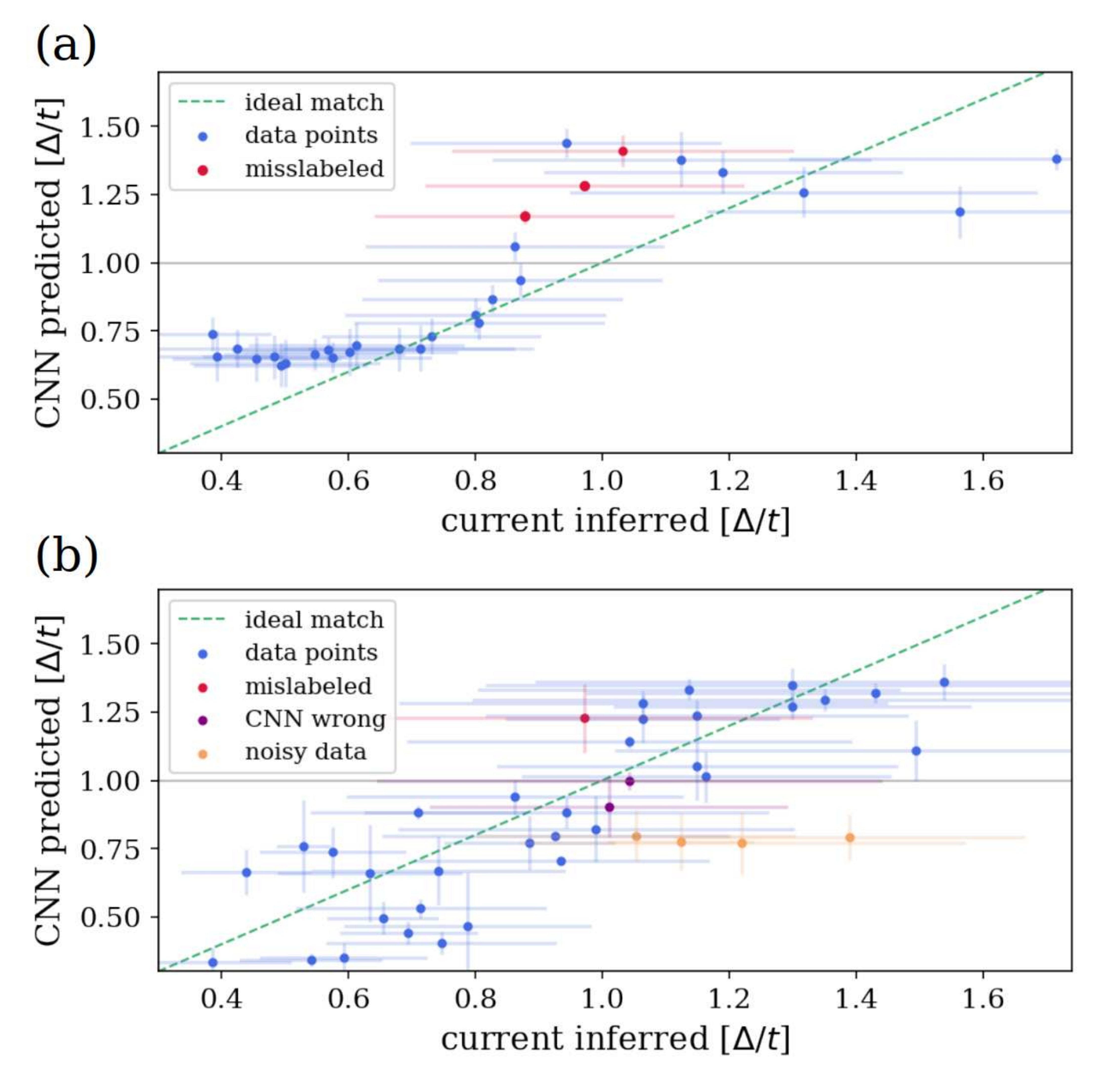}
    \caption{Predictions of the CNN (supervised) vs. experimental labels of the $\Delta/t$-ratio for the two experimental measurement series. The standard deviation of the prediction is given as an error bar for each data point. Panel (a) shows the 29 measurements at $B_\mathrm{Z} = \SI{150}{\milli\tesla}$ and (b) 39 measurements at $B_\mathrm{Z} = \SI{100}{\milli\tesla}$.
    } 
    \label{fig:predictions_CNN}
\end{figure}

Figure~\ref{fig:predictions_CNN} shows the results of the (supervised) CNN for the Hamiltonian inference process. Similar to Fig.~\ref{fig:predictions_cGAN}, Fig.~\ref{fig:predictions_CNN} shows the predictions of the CNN vs. current-inferred values of $\Delta/t$ for the first measurement series at $B_\mathrm{Z} = \SI{150}{\milli\tesla}$ in panel Fig.~\ref{fig:predictions_CNN}(a) and for the second at $B_\mathrm{Z} = \SI{100}{\milli\tesla}$ in panel Fig.~\ref{fig:predictions_CNN}(b).

For the first measurement series, we obtain with the CNN approach a mean error of $\delta_{CNN-1}=0.172$ and $\delta_{CNN-1}=0.150$ when neglecting the three mislabeled values. The classification accuracy, neglecting these three measurements is 92$\%$ (24/26 measurements). Compared to the predictions of the Conv-cGAN (see Fig.~\ref{fig:predictions_cGAN}), the CNN predictions show a step-function-like behavior and a plateau in the ECT and CAR phase, not following the "ideal match". 

The CNN predictions of the second series are shown in Fig.~\ref{fig:predictions_CNN}(b). Excluding the noisy measurements (orange dots) we obtain a mean error of $\delta_{CNN-2}=0.170$ and a classification accuracy of 94$\%$. As for the Conv-cGAN, the CNN predicts for two out of 35 measurements the wrong regime. The deviation from the "ideal match" is larger compared to the Conv-cGAN in Fig.~\ref{fig:predictions_cGAN}(b), especially for the ECT-dominated regime. However, we observe the general upwards trend, following the "ideal match" line mostly within the range of the experimental error bars.

\section{Experimental data pre-processing and workflow}
\label{App:workflow}
The workflow of the predictions for experimental data with the Conv-cGAN consists of five steps and is also shown in Figure~\ref{fig:results_exp} for two example measurements:
First, the original image (e.g. Fig.~\ref{fig:results_exp}(a,b)) of the differential conductance has to be cropped from 52$\times$52 to 28$\times$28 pixels (see Fig.~\ref{fig:results_exp}(c,d)) to match the dimension the Conv-cGAN is trained on.
Second, the cropped image has to be scaled according to the training data, which is normalized to values between 0 and 1 with an average maximum value of 0.88. Therefore, we normalize all experimental measurements to 0.88 to be consistent. 
Then, we define a parameter grid for the $\Delta/t$-ratio by splitting up the interval between 0.3 and 1.7 in 100 steps.
Afterward, the discriminator evaluates the cropped image with each point on the parameter grid, giving a "discriminator rating" for each $\Delta/t$-ratio yielding a probability distribution as shown in Fig.~\ref{fig:results_exp}(e,f) for two example measurements. The maximum value corresponds to the $\Delta/t$-ratio with the highest probability to belong to the corresponding differential conductance.
Finally, the first four steps are repeated 100 times for arbitrary croppings around the center position to obtain a standard deviation for the predictions of the discriminator network.\footnote{For some cases, the Conv-cGAN over- or underestimates the probability to exactly 0.3 or 1.7, respectively. This might be related to the under-sampling of the boundaries. Therefore, we exclude predictions of exactly 0.3 and 1.7 to obtain the standard deviation.}

\section{Extraction of the experimental labels}

\begin{figure}[h!]
    \centering
    \includegraphics[width=0.49\textwidth]{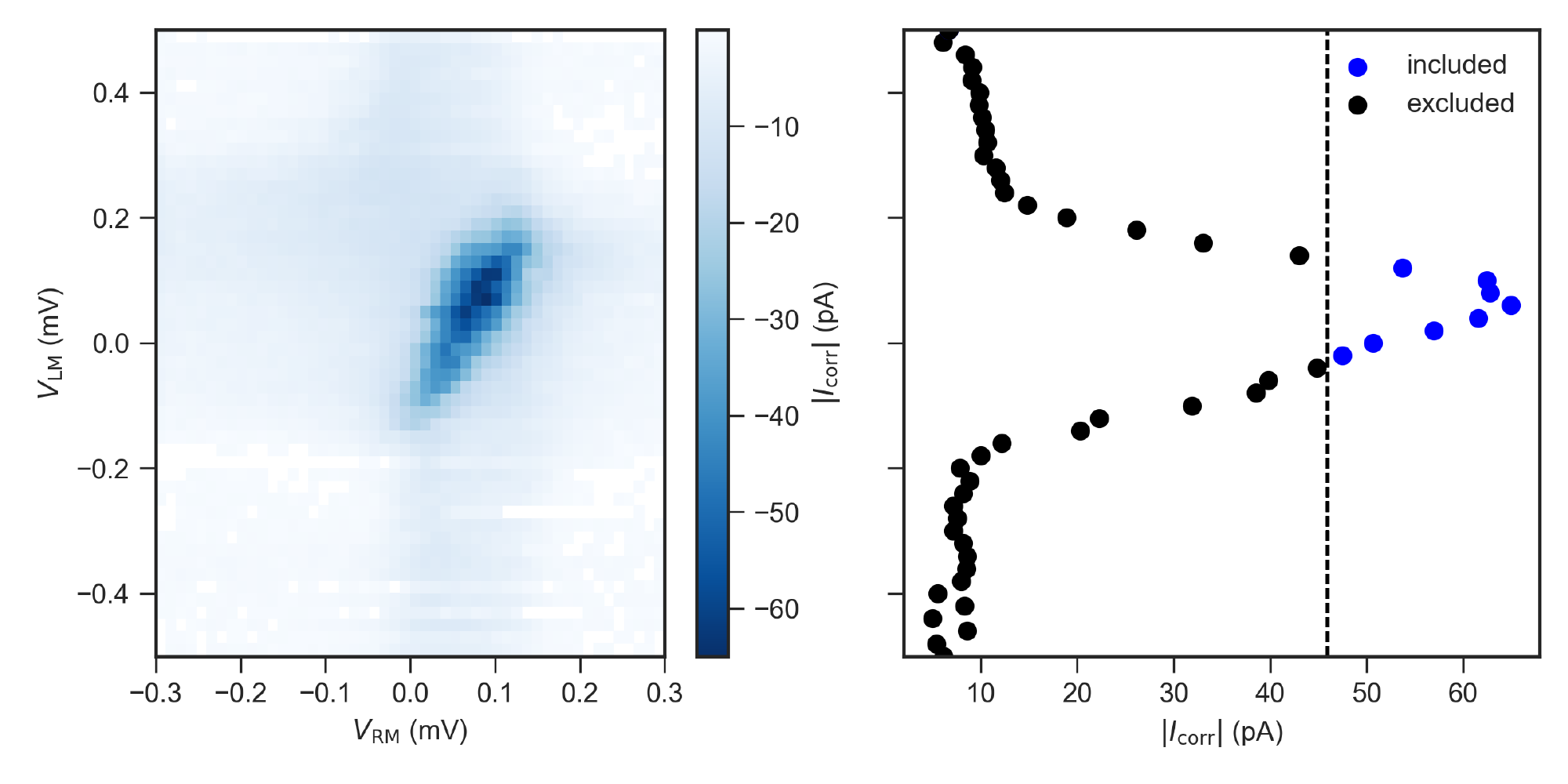}
    \caption{(left) The correlated current $I_\mathrm{corr} = \sqrt{-\mathrm{min}(I_\mathrm{L}\cdot I_\mathrm{R})}$ as measured on the left and right lead respectively, for varying QD plunger gates $V_\mathrm{LM}$ and $V_\mathrm{RM}$ at an anti-symmetric bias configuration 
    $V_\mathrm{L}=-V_\mathrm{R}=\SI{25}{\micro\electronvolt}$.
    (right) The maximum of $I_\mathrm{corr}$ for every linecut of $V_\mathrm{RM}$. The mean $\langle I_\mathrm{corr}\rangle$ and standard deviation $\sqrt{\langle I_\mathrm{corr}^2\rangle}$ are calculated from the points above the FWHM indicated by the black dashed line. $\langle I_\mathrm{corr}\rangle$ is taken to be $I_\mathrm{max}^\mathrm{ECT}$ due to the bias configuration allowing only ECT processes .}
    \label{fig:ECT_CAR_ratio_example}
\end{figure}

Figure \ref{fig:ECT_CAR_ratio_example} depicts a three-terminal measurement of correlated current as a function of the plunger gate voltages of two quantum dots arranged in a Kitaev chain configuration. The bias is set anti-symmetrically $V_\mathrm{L}=-V_\mathrm{R}$ to ensure that only elastic co-tunneling (ECT) can occur. The graph shows a positive slope indicating that current only arises when both dots are on resonance, indicative of ECT. By taking the maximum of $I_\mathrm{corr}$ for each value of $V_\mathrm{RM}$, we can estimate $I_\mathrm{max}^\mathrm{ECT}$ which is proportional to $t$ \cite{liu2022tunable}. 
Repeating this for a symmetric bias configuration $V_\mathrm{L}=V_\mathrm{R}$ yields $I_\mathrm{max}^\mathrm{CAR}$ which is proportional to $\Delta$. Then, a charge stability measurement is done for $V_\mathrm{L}=V_\mathrm{R}=0$, which shows the (un)-avoided crossings associated with CAR and ECT. These are labeled according to the $\Delta/t$ ratio obtained from the finite bias measurements.

\bibliography{references}

\end{document}